\documentclass[twocolumn,showpacs,aps,prb]{revtex4-1}

\usepackage{amssymb,amsmath}

\usepackage{graphicx}
\usepackage[caption=false]{subfig}
\usepackage[colorlinks,linkcolor=blue,citecolor=blue]{hyperref}

\begin{document}

 \title{Infinite-randomness criticality in a randomly layered Heisenberg magnet }

\author{Fawaz Hrahsheh}
\author{Hatem Barghathi}
%\affiliation{Department of Physics, Missouri University of Science and Technology, Rolla, MO 65409, USA}
\author{Thomas Vojta}
\affiliation{Department of Physics, Missouri University of Science and Technology, Rolla, MO 65409, USA}

\begin{abstract}
We study the ferromagnetic phase transition in a randomly layered Heisenberg magnet using large-scale Monte-Carlo
simulations. Our results provide numerical evidence for the infinite-randomness scenario recently predicted within 
a {strong-disorder renormalization group approach}. Specifically,
 we investigate the finite-size scaling behavior of the magnetic susceptibility which is characterized by
 a non-universal power-law divergence in the Griffiths phase.
 We also study the  perpendicular and parallel spin-wave stiffnesses in the Griffiths phase. In agreement with the theoretical 
predictions, the parallel stiffness is nonzero for all temperatures $T<T_c$. In contrast, the perpendicular stiffness 
remains zero in part of the ordered phase, giving rise to anomalous elasticity. In addition, we calculate the in-plane
 correlation length which diverges already inside the disordered phase at a temperature significantly higher than $T_c$.
The time autocorrelation function within model $A$ dynamics displays an ultraslow logarithmic decay at criticality and 
a nonuniversal power-law in the Griffiths phase.
\end{abstract}

\date{\today}

\maketitle

%%%%%%%%%%%%%%%%%%%%%%%%%%%%%%%%%%%%%%%%%%%%%%%%%%%%%%%%%%%%%%%%%%%%%%%%%%%%%%%%%
% Main text starts here
%%%%%%%%%%%%%%%%%%%%%%%%%%%%%%%%%%%%%%%%%%%%%%%%%%%%%%%%%%%%%%%%%%%%%%%%%%%%%%%%%
\section{Introduction}
%%%%%%%%%%%%%%%%%%%%%%%%%%%%%%%%%%%%%%%%%%%%%%%%%%%%%%%%%%%%%%%%%%%%%%%%%%%%%%%%%
When weak quenched disorder is added to a system undergoing a \emph{classical} continuous
phase transition, generically the critical behavior will either remain unchanged or it will be
replaced by another critical point with different exponent values. Which scenario is realized
depends on whether or not the clean critical point fulfills the Harris criterion.\cite{Harris74}
In contrast, zero-temperature quantum phase transitions generically display much stronger disorder
phenomena including power-law quantum Griffiths singularities,
\cite{ThillHuse95,GuoBhattHuse96,RiegerYoung96} infinite-randomness critical points
featuring exponential instead of power-law scaling, \cite{Fisher92,Fisher95}
and smeared phase transitions.\cite{Vojta03a,HoyosVojta08} A recent review of
these phenomena can be found in Ref.\ \onlinecite{Vojta06}, while Ref.\ \onlinecite{Vojta10}
focuses on metalic systems and also discusses experiments.

The reason for the disorder effects being stronger at quantum phase transitions than at
classical transitions is that quenched disorder is perfectly correlated in
the \emph{imaginary time} direction. Imaginary time behaves as an additional dimension at a
quantum phase transition and becomes infinitely extended at zero temperature.
Therefore, the impurities and defects are effectively ``infinitely  large'' in this extra dimension,
which makes them much harder to average out than the usual finite-size defects 
and so increases their influence.

For this reason, one should also expect strong unconventional disorder phenomena at classical
thermal phase transitions in systems in which the disorder is perfectly correlated in one or
more \emph{space} dimensions. Indeed, such behavior has been observed in the McCoy-Wu model, a
disordered classical two-dimensional Ising model having perfect disorder correlations
in one of the two dimensions. In a series of papers, McCoy and Wu \cite{McCoyWu68,McCoyWu68a,McCoyWu69,McCoy69}
showed that this model exhibits an unusual phase transition featuring a smooth specific heat
while the susceptibility is infinite over an entire temperature range.
Fisher \cite{Fisher92,Fisher95} achieved an essentially complete understanding of
this phase transition with the help of a strong-disorder renormalization group approach
(using the equivalence between the McCoy-Wu model and the random transverse-field Ising
chain). He determined that the critical point is of exotic infinite-randomness type
and is accompanied by power-law Griffiths singularities. In a classical Ising model with
perfect disorder correlations in \emph{two} dimensions, the disorder effects are even stronger 
than in the McCoy-Wu model: the sharp critical point is destroyed, and the transition is smeared
 over a range of temperatures.\cite{Vojta03b,SknepnekVojta04}

Recently, another classical system with perfect
disorder correlations in two dimensions was investigated by means of a strong-disorder
renormalization group.\cite{MohanNarayananVojta10} This theory predicts that the 
randomly layered Heisenberg magnet features a sharp critical point (in contrast to the Ising case discussed above).
However, it is of exotic infinite-randomness
type. Somewhat surprisingly, it is in the same universality class as the quantum critical point
of the random transverse-field Ising chain.

In this paper, we present the results of Monte-Carlo simulations of the randomly
layered Heisenberg model. They provide numerical evidence in support of the above
renormalization group predictions. Our paper is organized as follows. In Sec. \ref{section2},
we define our model and discuss its phase diagram. We also briefly summarize the predictions of the
strong disorder renormalization group theory.\cite{MohanNarayananVojta10} In Sec. \ref{section3}, we
describe our Monte-Carlo simulations, we present the results and compare them to the theory. We conclude in Sec.
\ref{section4}.

\section{Model and renormalization group predictions \label{section2}}
\label{sec:RG}

\begin{figure}
 \includegraphics[width=5cm,angle=-90, clip]{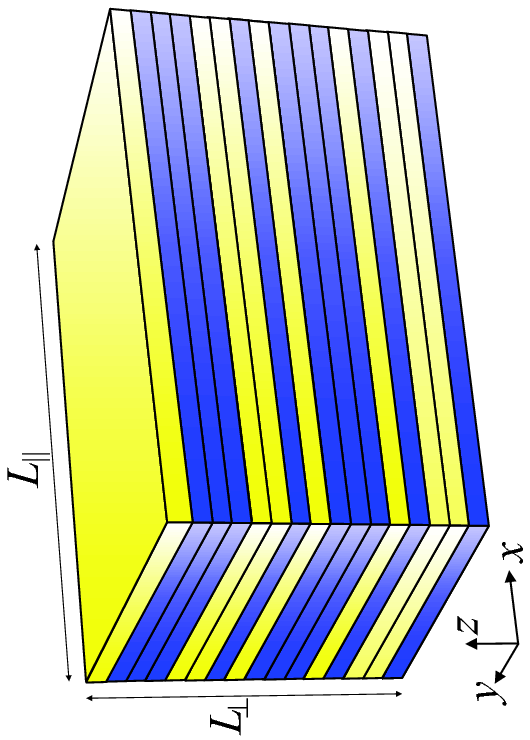}
\caption{(Color online) Schematic of the layered Heisenberg magnet:
 It consistes of a random sequence of layers of two different ferromagnetic materials.\cite{MohanNarayananVojta10}}
\label{lhmodel}
\end{figure}

We consider a ferromagnet consisting of a random sequence of
layers made up of two different ferromagnetic materials, see sketch in Fig.\
 \ref{lhmodel}.

 Its Hamiltonian, a classical Heisenberg
model on a three-dimensional lattice of perpendicular size $L_\perp$ (in $z$ direction) and
in-plane size $L_\parallel$ (in the $x$ and $y$ directions)  is given by
\begin{equation}
H = - \sum_{\mathbf{r}} J^{\parallel}_z \, (\mathbf{S}_{\mathbf{r}} \cdot \mathbf{S}_{\mathbf{r}+\hat{\mathbf{x}}}
                                        +\mathbf{S}_{\mathbf{r}} \cdot \mathbf{S}_{\mathbf{r}+\hat{\mathbf{y}}} )
    - \sum_{\mathbf{r}} J^{\perp}_z \, \mathbf{S}_{\mathbf{r}} \cdot\mathbf{S}_{\mathbf{r}+\hat{\mathbf{z}}}
    .
\label{Eq:Hamiltonian}
\end{equation}
Here, $\mathbf{S}_{\mathbf{r}}$ is a three-component unit vector on lattice site
$\mathbf{r}$, and  $\hat{\mathbf{x}}$, $\hat{\mathbf{y}}$, and $\hat{\mathbf{z}}$ are
the unit vectors in the coordinate directions. The interactions within
the layers, $J^{\parallel}_z$, and between the layers, $J^{\perp}_z$, are both positive
and independent random functions of the perpendicular coordinate $z$.

 In the following, we take all $J^{\perp}_z$ to be identical, $J^{\perp}_z \equiv J^{\perp}$,
while the $J^{\parallel}_z$ are drawn from a binary probability distribution
\begin{equation}
P(J^{\parallel})=(1-p)\, \delta(J^{\parallel} - J_u) + p\, \delta(J^{\parallel} - J_l) 
\label{BinaryDist}
\end{equation}
with $J_u > J_l$. Here, $p$ is the concentration of the ``weak'' layers while $1-p$ is the concentration of the ``strong'' layers.
 
\begin{figure}
 \includegraphics[width=8.5cm, clip]{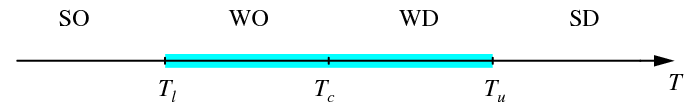}
\caption{(Color online) Schematic phase diagram of the randomly layered Heisenberg magnet
(\ref{Eq:Hamiltonian}). SD and SO denote the conventional strongly disordered
and strongly ordered phases, respectively. WD and WO are the weakly disordered
and ordered Griffiths phases. $T_c$ is the critical temperature while $T_u$ and $T_l$
mark the boundaries of the Griffiths phase .}
\label{Fig:pd}
\end{figure}

The qualitative behavior of the model (\ref{Eq:Hamiltonian}) is easily explained (see Fig. \ref{Fig:pd}). At sufficiently high
temperatures, the model is in a conventional paramagnetic (strongly disordered)
phase. Below a temperature $T_u$ (the transition temperature of a hypothetical system having
$J^{\parallel}_z \equiv J_u$ for all $z$) but above the actual critical temperature $T_c$,
rare thick slabs of strong layers develop local order while the bulk system is still
nonmagnetic. This is the \emph{paramagnetic} (weakly disordered) Griffiths phase (or Griffiths region). In the \emph{ferromagnetic} (weakly ordered) Griffiths phase,
located between $T_c$ and a temperature $T_l$ (the transition temperature of a hypothetical
system having $J^{\parallel}_z \equiv J_l$ for all $z$), bulk magnetism coexists with rare
nonmagnetic slabs.  Finally, below $T_l$, all slabs are locally ferromagnetic and the system is in a conventional ferromagnetic (strongly ordered) phase.

In Ref.\ \onlinecite{MohanNarayananVojta10}, the behavior in both Griffiths phases and at criticality
has been derived within a strong-disorder renormalization group calculation. Here, we simply
motivate and summarize the results. The probability of finding a slab of
$L_{RR}$ consecutive strong layers is given by simple combinatorics; it reads
$w(L_{RR}) \sim (1-p)^{L_{RR}} = e^{-\tilde p L_{RR} }$
with $\tilde p = -\ln(1-p)$. Each such slab is equivalent to a two-dimensional
Heisenberg model with an effective interaction $L_{RR} J_u$. Because the two-dimensional
Heisenberg model is exactly at its lower critical dimension, the renormalized
distance from criticality, $\epsilon$, of such a slab decreases exponentially with its thickness,
$\epsilon(L_{RR}) \sim e^{-b L_{RR}}$.\cite{Vojta06,VojtaSchmalian05}
Combining the two exponentials gives a power-law probability density of locally ordered slabs,
\begin{equation}
\rho(\epsilon) \sim \epsilon^{\tilde p / b -1} = \epsilon^{1/z-1}
\label{Eq:DOS}
\end{equation}
where the second equality defines the conventionally used dynamical exponent, $z$.
It increases with decreasing temperature throughout the Griffiths phase and
diverges as $z \sim 1/|T-T_c|$ at the actual critical point.

Many important observables follow from appropriate integrals of the density of states  (\ref{Eq:DOS}).
The susceptibility can be estimated by $\chi \sim \int d\epsilon\, \rho(\epsilon)/\epsilon$.
In an infinite system, the lower bound of the integral is 0; therefore, the susceptibility
diverges in the entire temperature region where $z>1$. A finite system size $L_\parallel$
in the in-plane directions introduces a nonzero lower bound $\epsilon_{\rm min} \sim
L_\parallel^{-2}$. Thus, for $z>1$, the susceptibility in the weakly disordered Griffiths phase diverges as
\begin{equation}
 \chi (L_\parallel) \sim L_\parallel^{2-2/z}
\label{Eq:Chi(Lt)WD}
\end{equation}
and in the weakly ordered Griffiths phase, it diverges as    
\begin{equation}
 \chi (L_\parallel) \sim L_\parallel^{2+2/z}.
\label{Eq:Chi(Lt)WO}
\end{equation}
 
The strong-disorder renormalization group \cite{MohanNarayananVojta10} confirms these simple estimates and gives
 $\chi\sim L_\parallel^2 [\ln{(L_\parallel/a)}]^{2\phi-1/\psi}$ at criticality where $\phi=(1+\sqrt{5})/2$ and $\psi=1/2$ 
are critical exponents of the infinite randomness critical point.

The spin-wave stiffness $\rho_s$ is defined by the work needed to twist the spins of two opposite boundaries by a relative 
angle $\theta$. Specifically, in the limit of small $\theta$ and large system size, the free-energy density $f$ depends on $\theta$ as
\begin{equation}
 f(\theta)-f(0)=\frac{1}{2}\rho_s\left(\frac{\theta}{L}\right)^2.
 \label{stiff}
\end{equation}
Because the randomly layered Heisenberg model is anisotropic, we need to distinguish the parallel spin-wave stiffness
 $\rho_s^\parallel$ from the perpendicular spin-wave stiffness $\rho_s^\perp$. To calculate the
parallel spin-wave stiffness, we apply boundary conditions at $x=0$ and $x=L_\parallel$ and set $L=L_\parallel$ in Eq. (\ref{stiff}) whereas the boundary
 conditions are applied at $z=0$ and $z=L_\perp$ to calculate the perpendicular spin-wave stiffness with $L=L_\perp$ in Eq. (\ref{stiff}).
  
Let us first discuss the parallel stiffness. In this case, the free energy difference $f(\theta)-f(0)$ is simply
the sum over all layers participating in the long-range order (each having the same twisted boundary conditions). 
Thus, $\rho_s^\parallel$ is nonzero everywhere in the ordered phase. The strong-disorder renormalization group approach
\cite{MohanNarayananVojta10} predicts
\begin{equation}
 \rho_s^\parallel \sim m \sim |T-T_c|^\beta~~~~~~~~~~~~~~(T<T_c)
\end{equation}
where $\beta=(3-\sqrt{5})/2$ is the order parameter exponent of the infinite randomness critical point.
The parallel stiffness behaves like the total magnetization $m=|\sum_{\mathbf{r}}{\langle \mathbf{S_r}\rangle}|/(L_\perp L_\parallel^2)$,
 because both renormalize additively under the strong-disorder renormalization-group theory.\cite{MohanNarayananVojta10}

If the twist $\theta$ is applied between the bottom ($z=0$) and the top ($z=L_\perp$) layers, the local
 twists between consecutive layers will vary from layer to layer. Minimizing $f(\theta)-f(0)$ leads to 
$\rho_s^\perp\sim \langle1/J_{eff}^\perp\rangle^{-1}$ where $J_{eff}^\perp$ are the effective couplings
between the rare regions. Within the strong-disorder renormalization group approach, the distribution
of the $J_{eff}^\perp$ follows a power law $p(J_{eff}^\perp)\sim (J_{eff}^\perp)^{1/z-1}$.
Thus, $\rho_s^\perp=0$ in part of the ordered Griffiths phase. It only becomes nonzero once $z$ falls below $1$
at a temperature $T_s<T_c$. Between $T_c$ and $T_s$, the system displays 
anomalous elasticity. Here, the free energy due to the twist scales with $f(\theta)-f(0) \sim L_\perp^{-1-z}$.
Thus, the perpendicular stiffness formally vanishes as $\rho_s^\perp\sim L_\perp^{1-z}$ with increasing $ L_\perp$.

To study the dynamical critical behavior, a phenomenological dynamics is added to the randomly layered Heisenberg model.
 The simplest case is a purely relaxational dynamics corresponding to model $A$ in the classification of 
Hohenberg and Halperin.\cite{HohenbergHalperin77}

The dynamic behavior can be characterized by the average time autocorrelation function
\begin{equation}
 C(t)=\frac{1}{L_\perp L_\parallel^2}\int{d^3r \langle\mathbf{S}_{\mathbf{r}}(t)\mathbf{S}_{\mathbf{r}}(0)\rangle},
\end{equation}
where $\mathbf{S_r}(t)$ is the value of the spin at position $\mathbf{r}$ and time $t$.

The behavior of $C(t)$ in the weakly disordered Griffiths phase can be easily estimated. The correlation time of a single locally ordered 
slab is proportional to $1/\epsilon$.\cite{MohanNarayananVojta10} Summing over all slabs using the density of states
(\ref{Eq:DOS}) then gives
\begin{equation}
 C(t)\sim \int{d\epsilon \rho(\epsilon)e^{-\epsilon t}}\sim t^{-1/z}.
\label{Ctvst}
\end{equation}

The strong disorder renormalization group calculation \cite{MohanNarayananVojta10} confirms this estimate. 
Moreover, at criticality, when $z\to \infty$, it gives an even slower logarithmic behavior
\begin{equation}
C(t) \sim [\ln(t/t_0)]^{\phi-1/\psi}.
\label{eq:C_CP}
\end{equation}
where $t_0$ is a microscopic length scale.

\section{Monte-Carlo simulations \label{section3}}
\subsection{Overview}
In this section we report  results of Monte-Carlo simulations of the randomly
layered Heisenberg magnet. Because the phase transition in this system is dominated by the
rare regions, sufficiently large system sizes are required in  order to get reliable results. We have simulated system
sizes ranging from $L_\perp=90$ to $800$ and $L_\parallel=10$ to $400$. We have chosen $J_u=1$ and $J_l=0.25$ in Eq.\ ({\ref{BinaryDist}}).
All the simulations have been performed for disorder concentrations $p=0.8$. With these parameter choices, the Griffiths 
region ranges from $T_l\approx0.63$ to $T_u\approx1.443$. For optimal performance,
 we have used large numbers of disorder realizations, ranging from $100$ to $7200$,
 depending on the system size. While studying the thermodynamics,
we have used the efficient Wolff cluster algorithm \cite{Wolff89} to eliminate critical slowing down.
 We have equilibrated every run by 100 Monte-Carlo sweeps, and we have used another 100 sweeps for measurements.
 To investigate the critical dynamics,
 we have equilibrated the system using the Wolff algorithm but then 
propagated the system in time by means of the Metropolis algorithm \cite{MRRT53} which implements model $A$ dynamics. 

\subsection{Thermodynamics \label{subsection3a}}
\begin{figure}
\includegraphics[width=8.1cm]{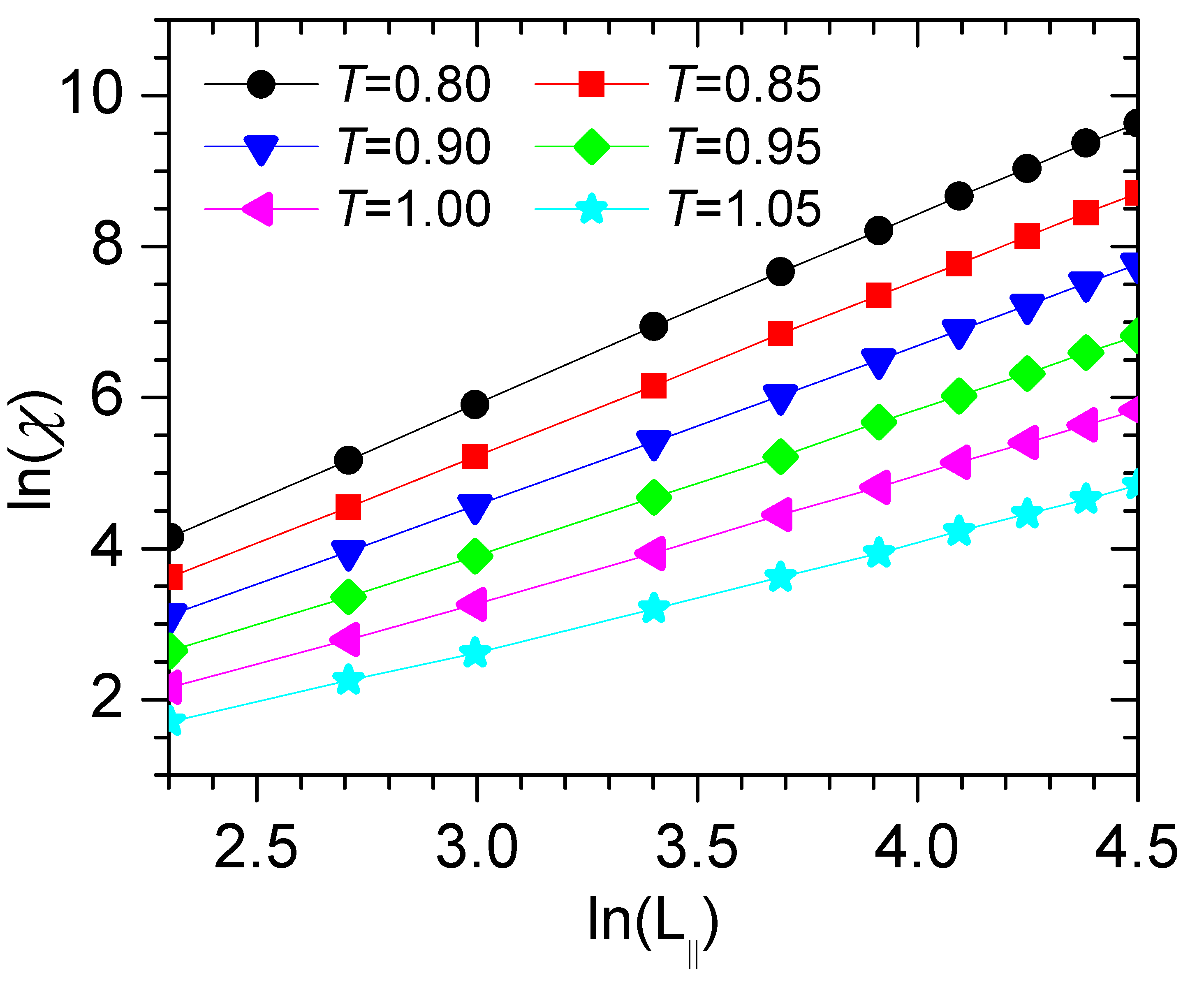}
\caption{(Color online) Susceptibility $\chi$ as a function of in-plane system size $L_\parallel$
for several temperatures in the Griffiths region. The perpendicular size is $L_\perp=800$;
the data are averages over 300 disorder configurations. The solid lines are fits
to the power laws (\ref{Eq:Chi(Lt)WD}, \ref{Eq:Chi(Lt)WO}).}
\label{lnchivslnLII}
\end{figure}

To test the finite-size behavior (\ref{Eq:Chi(Lt)WD}, \ref{Eq:Chi(Lt)WO}) of the susceptibility, one needs to consider samples having sizes
$L_\perp \gg L_\parallel$ such that $L_\perp$ is effectively infinite. We have used system sizes
$L_\perp=800$ and $L_\parallel=10$ to $90$. Figure \ref{lnchivslnLII} shows the susceptibility $\chi$ as a function of $L_\parallel$
for several temperatures in the Griffiths region between $T_l=0.63$ and $T_u \approx 1.443$. In agreement with the theoretical
predictions (\ref{Eq:Chi(Lt)WD}) and (\ref{Eq:Chi(Lt)WO}),
$\chi$ follows a nonuniversal power law
in  $L_\parallel$ with a temperature-dependent exponent. Simulations for many more
temperature values, in the range $T\approx0.76-1.2$, yield analogous results.

\begin{figure}
\includegraphics[width=8cm]{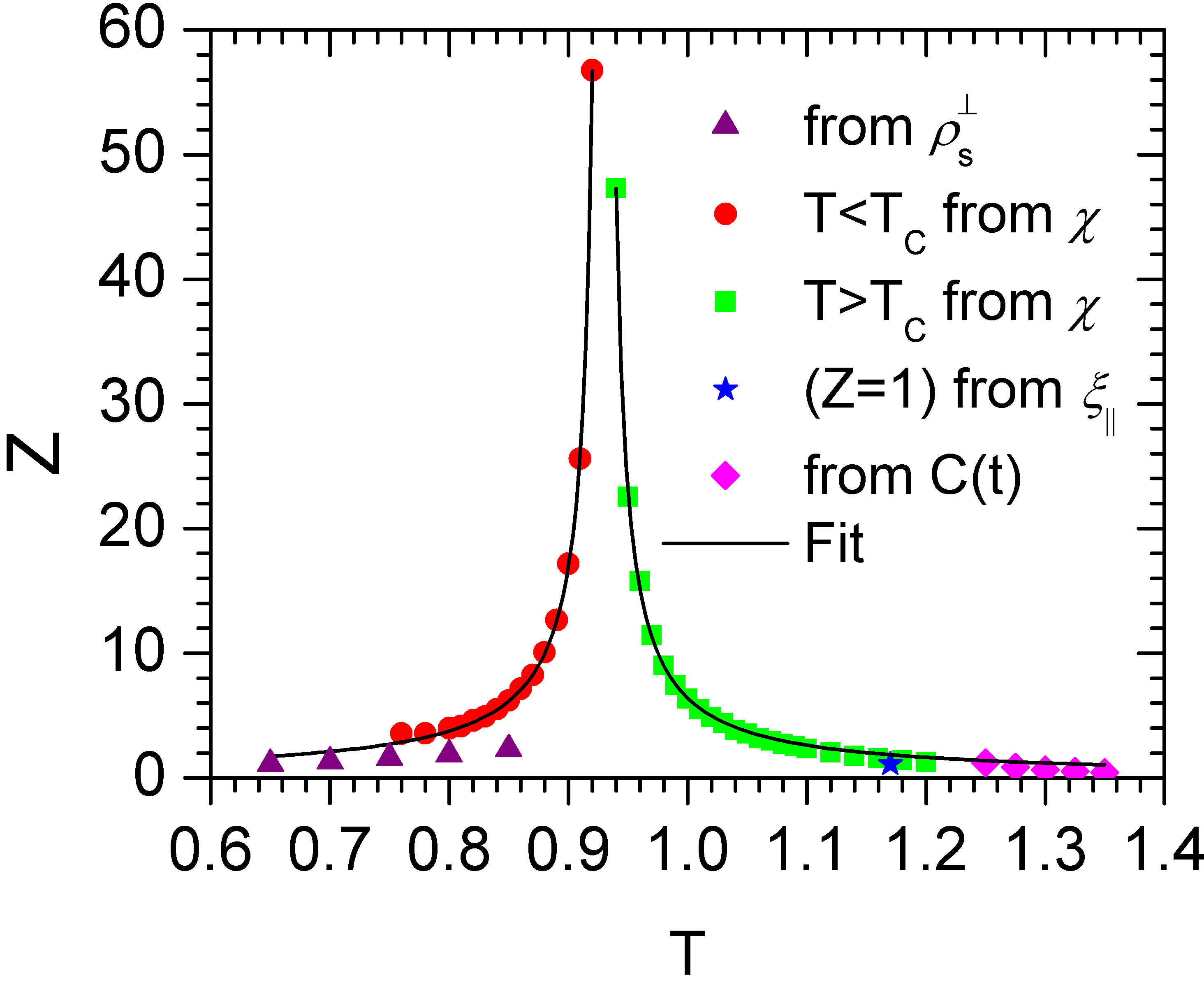}
\caption{ (Color online) Griffiths dynamical exponent $z$ vs temperature. The data are extracted from the perpendicular stiffness data in Fig. \ref{lnrhosvslnL},
the susceptibility data in Fig. \ref{lnchivslnLII}, the parallel correlation length data in Fig. \ref{xiIIbyLIIvsT} 
and the autocorrelation function data in Fig. \ref{lnCtvslnt}. The solid lines are a power-law fit of
 $z$ (extracted from Fig. \ref{lnchivslnLII}) to (\ref{Eq:Chi(Lt)WD}) and (\ref{Eq:Chi(Lt)WO}). }
\label{zvsT}
\end{figure}

 The values of the exponent $z$ extracted from fits to
(\ref{Eq:Chi(Lt)WD}, \ref{Eq:Chi(Lt)WO}) are shown in Fig. \ref{zvsT} for the \textit{paramagnetic} and \textit{ferromagnetic} sides of
the Griffiths region. $z$ can be fitted to the predicted power law $z\sim 1/|T-T_c|$, as discussed
after (\ref{Eq:DOS}), giving the estimate $T_c \approx 0.933$.
 
\begin{figure}
\includegraphics[width=8.5cm]{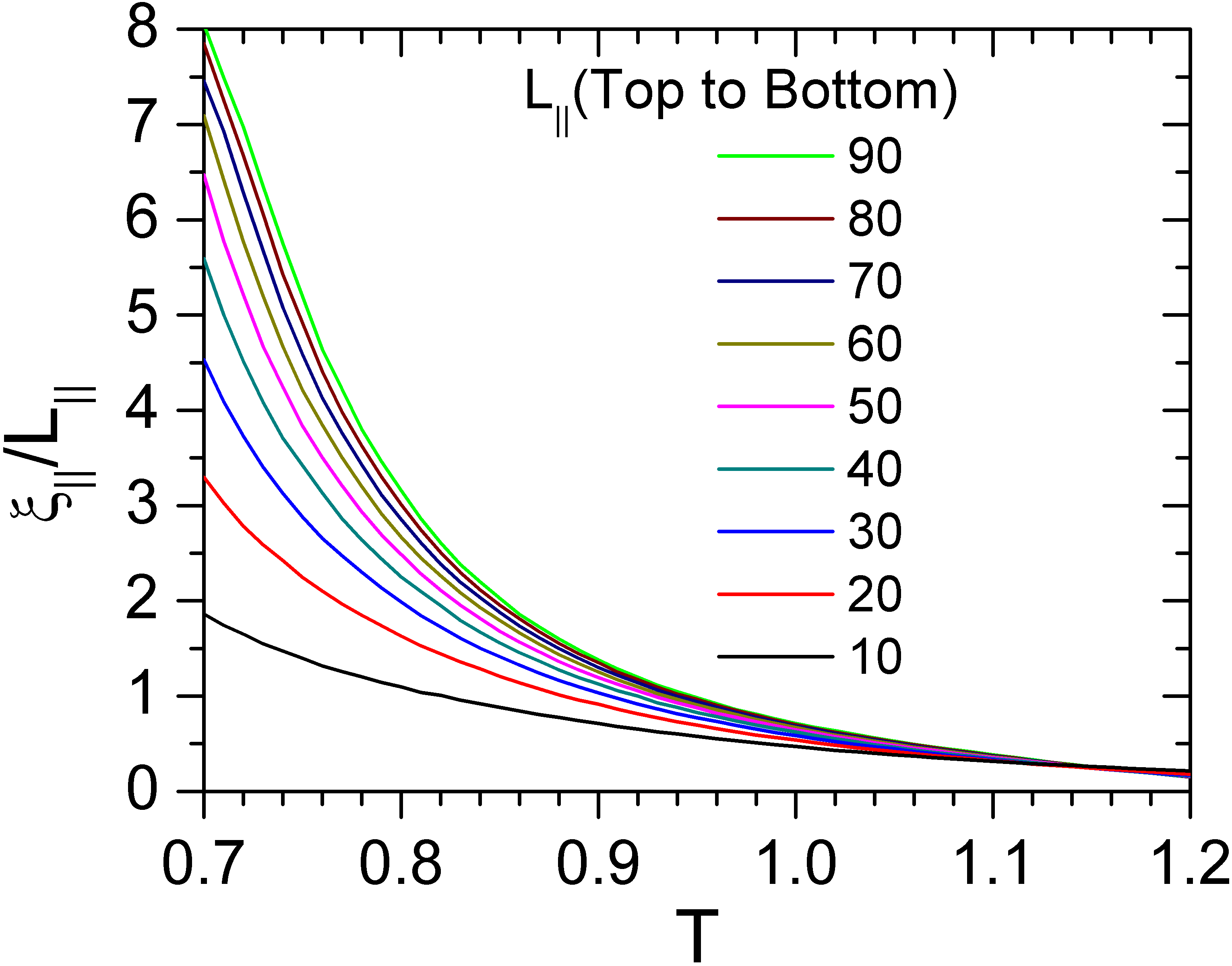}
\caption{ (Color online) Scaled in-plane correlation length $\xi_\parallel/L_\parallel$ as a function of temperature $T$
         for several in-plane system sizes $L_\parallel$ in the Griffiths region. The perpendicular size is $L_\perp=800$;
         the data are averaged over 300 disorder configurations.}
\label{xiIIbyLIIvsT}
\end{figure}

For a deeper understanding  of the thermodynamic critical phenomena of the layered Heisenberg model, we have also
studied the behavior of the in-plane correlation lengths in Griffiths phase. Figure \ref{xiIIbyLIIvsT} shows the scaled correlation
length $\xi_\parallel/L_\parallel$ as a function of temperature for different values of $L_\parallel$.
Surprisingly, the curves cross at a temperature, $T\approx 1.17$, significantly higher than $T_c \approx 0.93$.
 This implies that the average in-plane correlation length diverges in part of the \emph{disordered} phase.

To understand this behavior, we estimate the rare region contribution to the averaged in-plane correlation
length. It can be calculated by integrating over the density of states (\ref{Eq:DOS}) as

\begin{equation}
 \xi_\parallel^2\sim \int_0^{\epsilon_0}{d\epsilon \rho(\epsilon) \xi_\parallel^2(\epsilon)}\sim \int_0^{\epsilon_0}{d\epsilon\epsilon^{1/z-1}\frac{1}{\epsilon}}
\label{xi}
\end{equation}
where $ \xi_\parallel^2(\epsilon)\sim 1/\epsilon$ is the dependence of the in-plane correlation length of a single region
 \cite{MohanNarayananVojta10,Bray88}
 on the renormalized distance $\epsilon$ from criticality. Note that we average $\xi_\parallel^2$ instead 
of $\xi_\parallel$ because that is what numerically happens in the \textit{second moment method} which defines $\xi_\parallel^2$ via

\begin{equation}
 \xi_\parallel^2=\frac{\sum_{\mathbf r}{C(\mathbf r) \mathbf r^2}}{\sum_{\mathbf r}{C(\mathbf r)}}
\end{equation}
with $C(\mathbf r)$ being the spatial correlation function. The integral in (\ref{xi}) diverges for $z>1$ and converges for $z<1$.
The in-plane correlation length therefore diverges already in the disordered Griffiths phase at the temperature
at which the Griffiths dynamical exponent is $z=1$. From Fig. \ref{xiIIbyLIIvsT} we estimate this temperature
 to be $T\approx1.17$. As can be seen in Fig. \ref{zvsT}, this value is in good agreement with the result extracted from the finite 
size behavior of $\chi$.

 We now turn to the spin-wave stiffness. Calculating the stiffness by actually carrying out simulations with twisted boundary conditions is not very efficient.
 However, the stiffness can be rewritten in terms of expectation values calculated in a conventional run with periodic 
boundary conditions. The resulting formula which is a generalization of that used by Caffarel \emph{et al} \cite{CaffarelAzariaDelamotteMouhanna94}
reads
\begin{equation}
\begin{gathered}
 \rho_s^\perp=\left\langle\sum_{\langle\mathbf r,\mathbf r'\rangle}J_{\mathbf r, \mathbf r'}\left[\mathbf S_{\mathbf r}\cdot\mathbf S_{\mathbf r'}-
(\mathbf S_{\mathbf r}\cdot \hat {\mathbf a})(\mathbf S_{\mathbf r'}\cdot \hat {\mathbf a})\right](z-z')^2 \right\rangle\\
-\frac{1}{T}\left\langle\left(\sum_{\langle\mathbf r,\mathbf r'\rangle}J_{\mathbf r, \mathbf r'}\left[(\mathbf S_{\mathbf r}\times \mathbf S_{\mathbf r'})\cdot
 \hat {\mathbf a}\right](z-z')\right)^2\right\rangle.
\end{gathered}
\end{equation}

 Here, $\hat {\mathbf a}$ can be any unit vector perpendicular to the total magnetization $\mathbf m$.
 For $\rho_s^\parallel$, $(z-z')$ has to be replaced by $(x-x')$.
 This formula is derived in appendix A.

Figure \ref{rhovsT} shows the results for the perpendicular and parallel stiffnesses of our randomly layered Heisenberg model.
 We have used a system of size $L_\perp=100$ and $L_\parallel=400$. The figure shows that the two stiffness indeed behave
very differently. The parallel stiffness $\rho_s^\parallel$ vanishes at $T\approx0.9-0.95$ in good agreement with our earlier estimate of 
$T_c\approx0.93$. In contrast, the perpendicular stiffness vanishes at a much lower temperature $T\approx 0.7$.
 Thus, in the range between $T\approx 0.7$ and $T_c$, the system displays anomalous elasticity, as predicted. (Note:
 The slight rounding of both $\rho_s^\parallel$ and $\rho_s^\perp$ can be attributed to finite-size effects.)

The results of the perpendicular spin-wave stiffness $\rho_s^\perp$ are analyzed in more detail in Fig. \ref{lnrhosvslnL} for perpendicular sizes $L_\perp=15-40$.
 We have used a parallel size $L_\parallel=400$ and a temperature range $T=0.65-0.85$ where the data are averaged over 1000 disorder configurations.
 The plot shows a non-universal power-law dependence of $\rho_s^\perp$ on $L_\perp$ which agrees with the prediction 
\begin{equation}
  \rho_s^\perp\sim L_\perp^{1-z}.
\label{rhosofL}
\end{equation}
 The dynamical exponents $z$ extracted from fits of $\rho_s^\perp$ to (\ref{rhosofL})
are also shown in Fig. \ref{zvsT}. While they roughly agree with the values extracted from $\chi$,
 the agreement is not very good. We believe this is due to the rather small $L_\perp$ values used.
 
\begin{figure}
\subfloat{\label{rhovsT}\includegraphics[width=8.5cm]{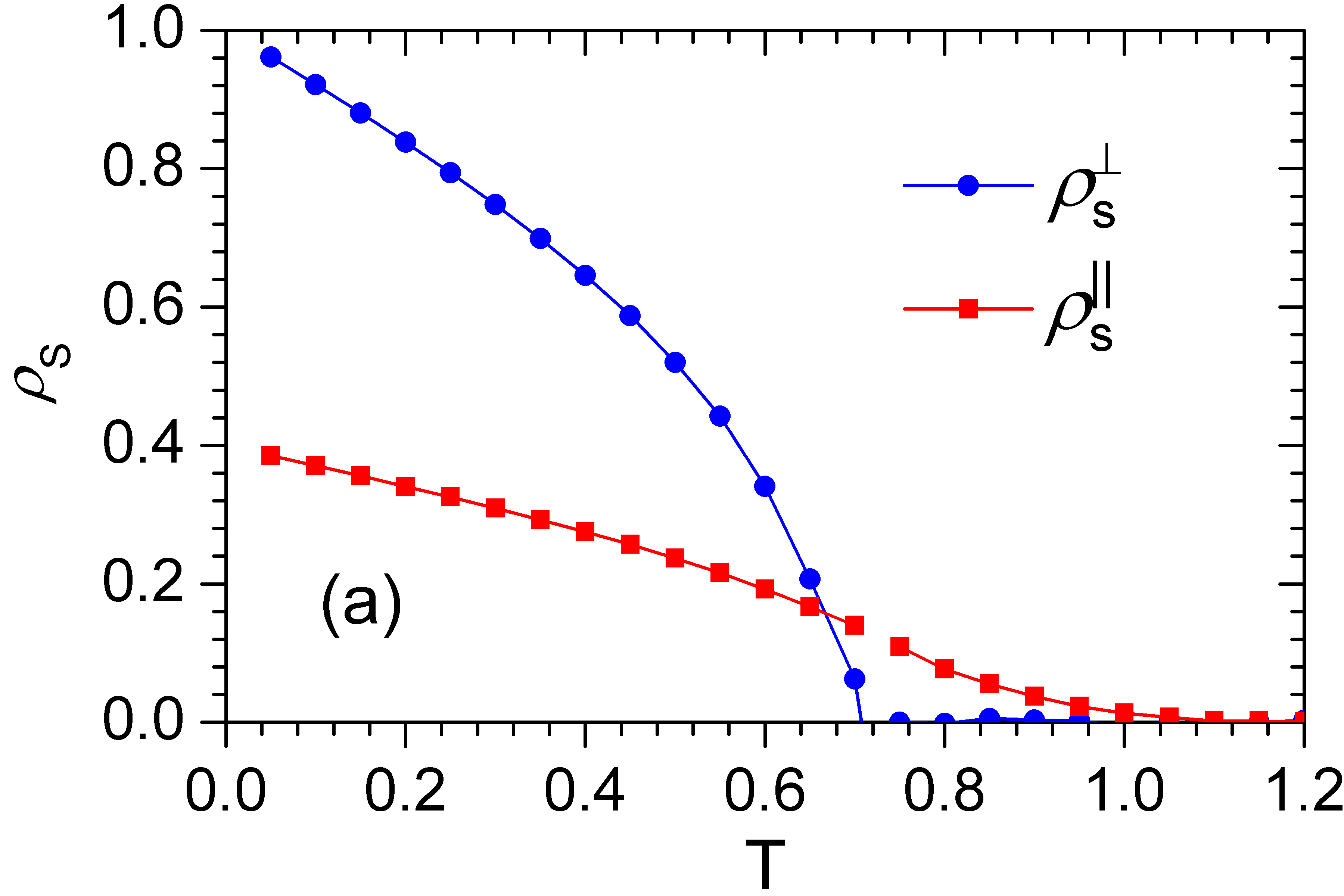}}\\
\subfloat{\label{lnrhosvslnL}\includegraphics[width=8.5cm]{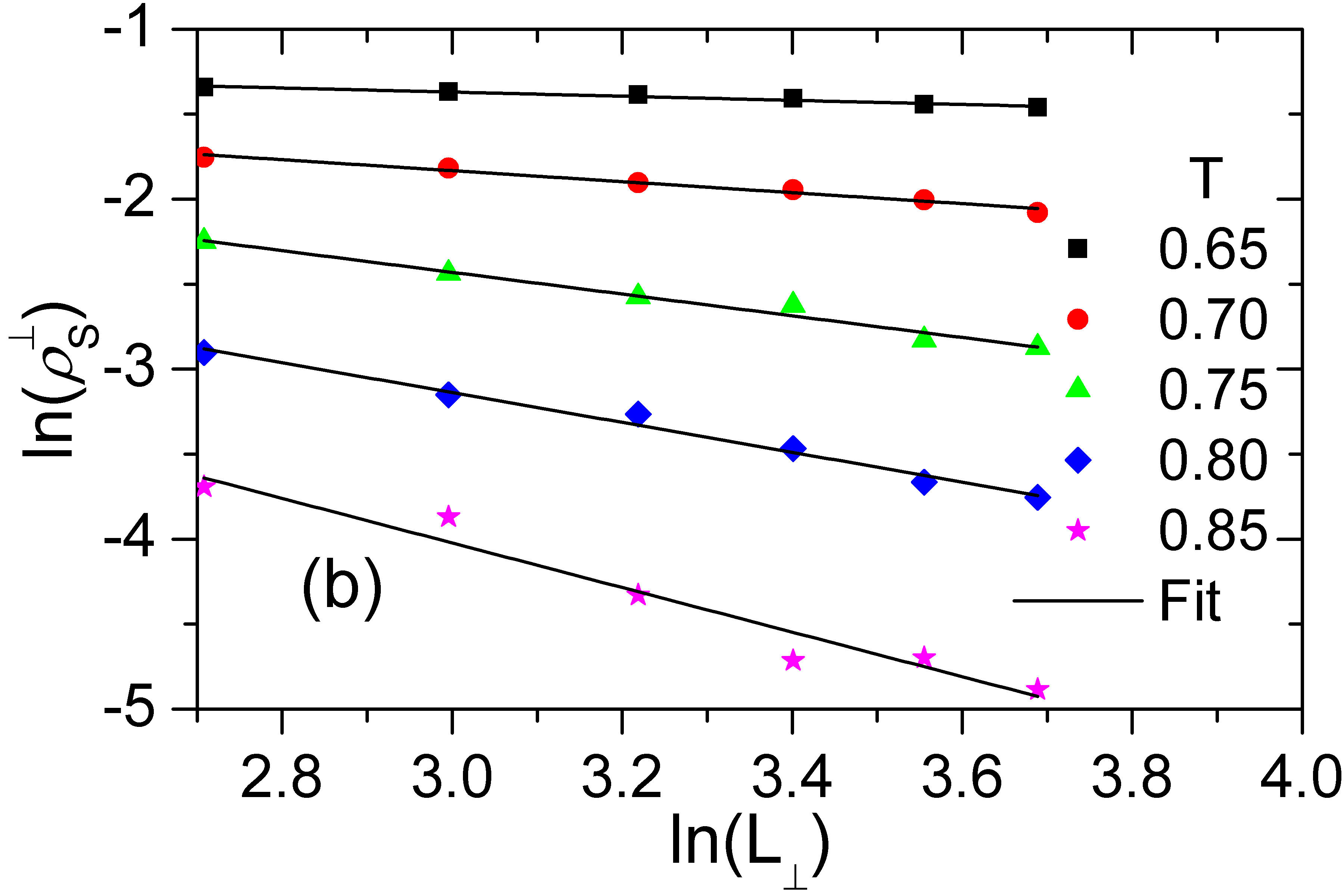} }
 \caption{(Color online) a: Perpendicular and parallel spin-wave stiffnesses
  ($\rho_s^\perp$ and $\rho_s^\parallel$, respectively) as functions of temperature $T$ 
 for system with sizes $L_\perp=100$ and $L_\parallel=400$. The data are averaged over 100 disorder configurations.
b: Perpendicular spin-wave stiffness as a function of $L_\perp$ for temperatures in
the weakly ordered Griffiths phase and $L_\parallel=400$. The data are averaged over 1000 disorder configurations.
The solid lines are fits to (\ref{rhosofL}).}
 \end{figure}
 
 \subsection{Critical dynamics \label{subsection3b}}
\begin{figure} 
\includegraphics[width=8cm]{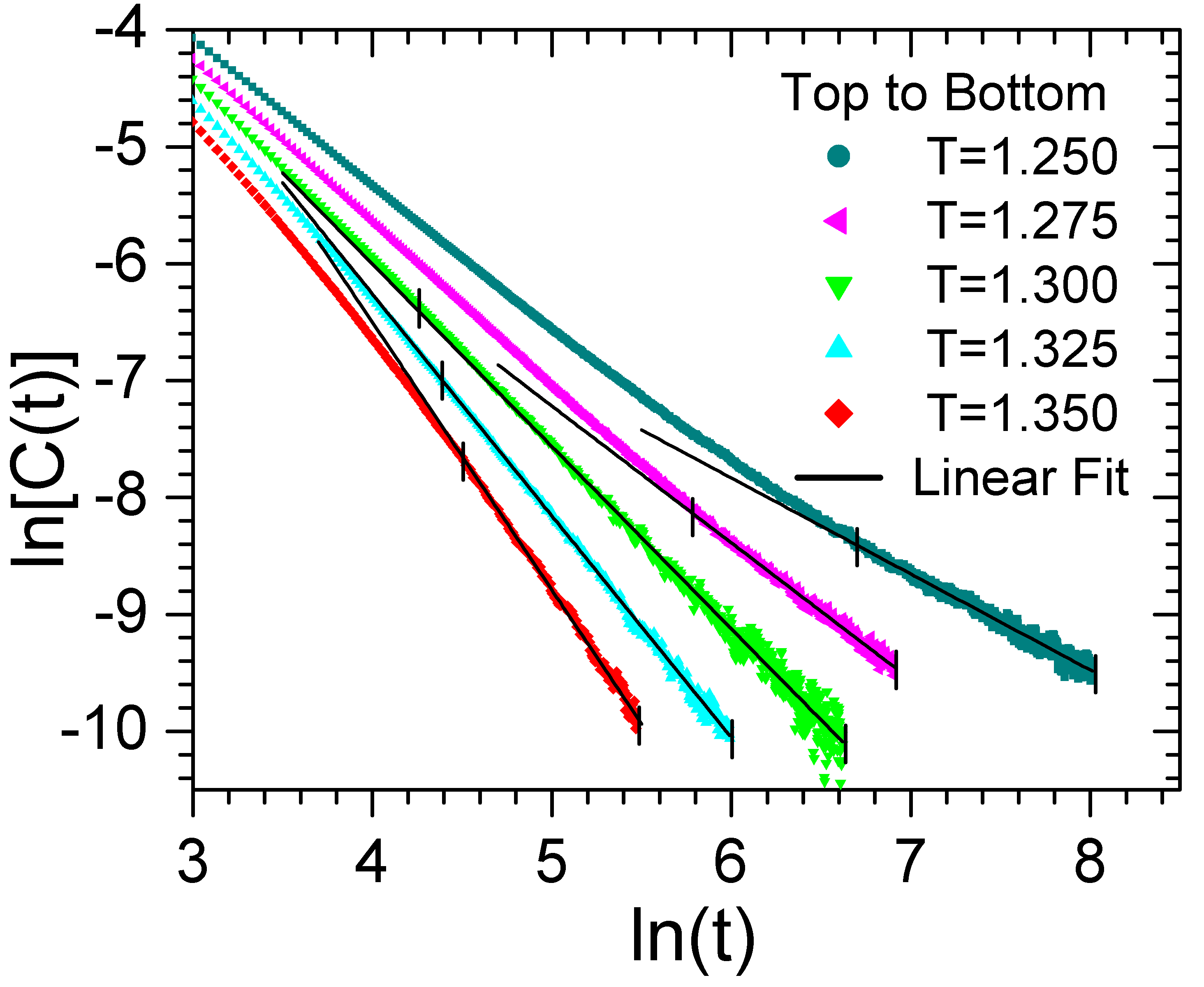}
\caption{ (Color online) Time autocorrelation function $C(t)$ for temperatures from $T=1.25$ to $1.35$ (within the Griffiths phase).
The system sizes are $L_\perp=400$ and $L_\parallel=100$. The data are averaged over $1720-7200$ disorder configurations.
The solid lines are fits to the power-law prediction (\ref{Ctvst}) (with the fit range marked).}
\label{lnCtvslnt}
\end{figure}

To investigate the behavior of the autocorrelation function $C(t)$ in the weakly disordered Griffiths phase, we have used 
system sizes $L_\perp=400$ and $L_\parallel=100$ and temperatures from $T=1.25$ to $1.35$. From figure \ref{lnCtvslnt}, one can 
see that the long-time behavior of $C(t)$ in the Griffiths phase follows a non-universal power law which is in agreement with the prediction (\ref{Ctvst}).
Fits of the data to (\ref{Ctvst}) can be used to obtain yet another estimate of the dynamical exponent $z$. The resulting
values are shown in Fig. \ref{zvsT}, they are in good agreement with those extracted from $\chi$.

\begin{figure} 
\includegraphics[width=8cm]{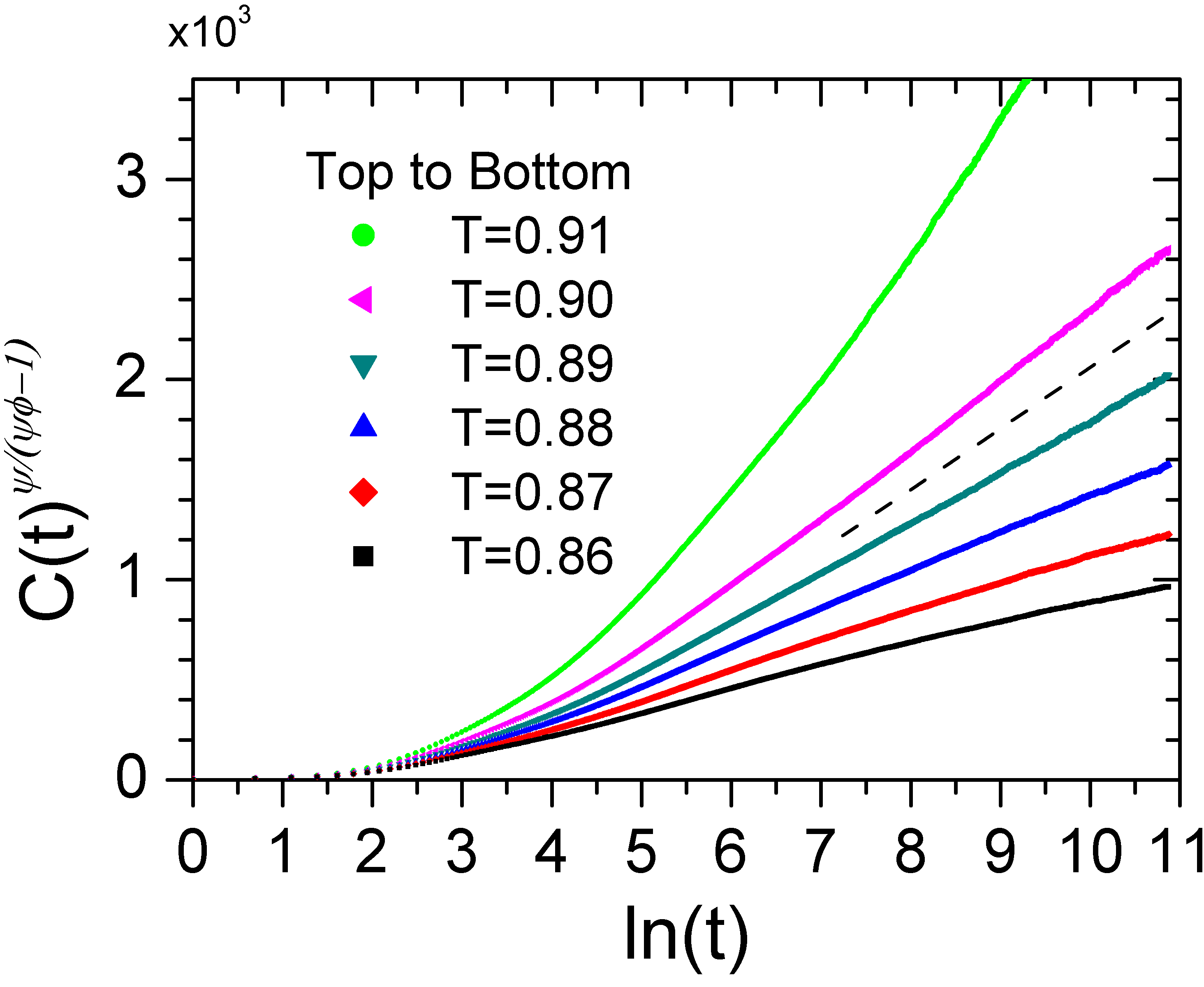}
\caption{ (Color online) Time autocorrelation function $C(t)$ for temperatures from $T=0.86$ to $0.91$ (near criticality).
The system sizes are $L_\perp=400$ and $L_\parallel=230$. The data are averaged over 70 to 80 disorder configurations.
The dashed line shows the logarithmic behavior (\ref{eq:C_CP}) at the estimated critical temperature $T_c=0.895$. }
\label{Ctvslnt}
\end{figure}

Figure \ref{Ctvslnt} shows the behavior of $C(t)$ near criticality plotted such that the expected logarithmic time-dependence
(\ref{eq:C_CP}) gives a straight line. We have used system sizes $L_\perp=400$ and $L_\parallel=230$
and temperatures from $T=0.86$ to $0.91$. We find that $C(t)$ indeed follows the prediction at an estimated $T_c\approx0.895$.
This estimate agrees reasonably well with that stemming from the finite-size behavior of $\chi$. We attribute the 
remaining difference to the finite-size effects and (in case of $C(t)$) finite-time effects.

\section{Conclusions \label{section4}}
To summarize, we have reported the results of large-scale Monte-Carlo simulations of the thermodynamics and
dynamic behavior of a randomly layered Heisenberg model. Our results provide strong numerical evidence in support
of the infinite-randomness scenario predicted within the strong-disorder renormalization group approach.\cite{MohanNarayananVojta10}
Morever, our data are compatible with the prediction that the randomly layered Heisenberg model is in the same universality
class as the one-dimensional random transverse-field Ising model.

We would have liked to determine the complete set of critical exponents of the infinite-randomness critical point directly from the numerical data.
To this end we have attempted to perform an anisotropic finite-size scaling analysis as in Refs. \onlinecite{PYRK98} or 
\onlinecite{SknepnekVojtaVojta04}. However, within the accessible range of system sizes of up to about $10^7$ sites, the 
corrections to the leading scaling behavior were so strong that we could not complete the analysis. This task thus remains for the future.

An important question left unanswered by the strong-disorder renormalization group approach\cite{MohanNarayananVojta10} is
whether or not weakly or moderately disordered systems actually flow to the infinite-randomness critical point. The 
clean Heisenberg critical point is unstable against weak layered disorder because it violates the generalized Harris criterion $d_r\nu>2$
where $d_r=1$ is the number of random dimensions. Thus, weak layered randomness initially increases under renormalization.
Our numerical parameter choices, $p=0.8$ and $J_u/J_l=4$ correspond to moderate disorder as the distribution is not 
particularly broad \emph{on a logarithmic scale}. The fact that we do confirm infinte-randomness behavior for these parameters
suggests that the infinite-randomness critical point may control the transition for any nonzero disorder strength.
A numerical verification of this conjecture by simulating very weakly disordered systems would require even larger
system sizes and is thus beyond our present computational capabilities.

Experimental verifications of infinite-randomness critical behavior and the
accompanying power-law Griffiths singularities have been hard to come
by, in particular in higher-dimensional systems. Only very recently, promising
measurements have been reported \cite{Westerkampetal09,UbaidKassisVojtaSchroeder10}
of the quantum phase transitions in CePd$_{1-x}$Rh$_x$ and Ni$_{1-x}$V$_x$.
The randomly layered Heisenberg magnet considered here provides an alternative realization
of an infinite-randomness critical point.  It may be more easily realizable in experiment
because the critical point is classical, and samples can be produced by depositing
random layers of two different ferromagnetic materials.

Magnetic multilayers with systematic variation of the critical temperature from layer to layer have 
already been produced,\cite{MPHW09} and our results would apply to random versions of these structures.

\section*{Acknowledgements}
We acknowledge helpful discussions with S. Bharadwaj, P. Mohan, and R. Narayanan. 
This work was supported in part by the NSF under grant No. DMR-0906566.

\appendix
\section{Spin-wave stiffness in terms of spin correlation functions \label{appendixa}} 
Twisted boundary conditions, i.e., forcing the spins on one surface of the sample of size $L$ to make an angle of $\theta$
with those on the opposite surface, lead to a change in the free energy density $f$. It can be parametrized by
\begin{equation}
 f(\theta)-f(0)=\frac{1}{2} \rho_s \left(\frac{\theta}{L}\right)^2.
\label{deltaf}
\end{equation} 
which defines the spin-wave stiffness $\rho_s$.

For definiteness, assume we apply a twist of $\theta$ around the perpendicular axis between the top and bottom
surfaces of the sample.
We parametrize the Heisenberg spin as
\begin{equation}
 \mathbf{S}_{\mathbf r} = \begin{pmatrix} \sin(\vartheta_{\mathbf r}) \cos(\phi_{\mathbf r})\\
 \sin(\vartheta_{\mathbf r}) \sin(\phi_{\mathbf r})\\ \cos(\vartheta_{\mathbf r}) \end{pmatrix}.
\label{spinform}
\end{equation}
The boundary conditions then read $\phi_{\mathbf r}=0$ at the bottom $(z=0)$ surface and 
$\phi_{\mathbf r}=\theta$ at the top $(z=L_\perp)$ surface. To eliminate the twisted boundary condition,
 we now perform the variable transformation
\begin{equation}
 \psi_{\mathbf r}=\phi_{\mathbf r}-\theta\frac{z_{\mathbf r}}{L_\perp}
  \label{psi}
\end{equation}
which gives new boundary conditions of $\psi_{\mathbf r}=0$ at both $z_{\mathbf r}=0$ and $z_{\mathbf r}=L_\perp$.

Substituting the variable transformation in the Heisenberg Hamiltonian (\ref{Eq:Hamiltonian}), we obtain 
\begin{equation}\begin{split}\label{VecHamiltonian}
H  =& -\sum_{\langle\mathbf r,\mathbf r'\rangle} {J_{\mathbf r,\mathbf r'}} \bigg\{ \sin(\vartheta_{\mathbf r})\ \sin(\vartheta_{\mathbf r'}) \\
& \cos{ \bigg(\psi_{\mathbf r}-\psi_{\mathbf r'}+\frac{\theta}{L_\perp}(z-z') \bigg)}+ \cos(\vartheta_{\mathbf r})\ \cos(\vartheta_{\mathbf r'})\bigg\}\\
\end{split}\end{equation}
where the twist is ``distributed'' over the volume.
Thus, the twist angle $\theta$ now appears as a parameter of the Hamiltonian. We can
use standard methods to reformulate the second derivative of the free energy $F$ as
\begin{equation}
  \frac{\partial ^2 F}{\partial \theta^2}=\frac{1}{T} \left\langle\frac{\partial H}{\partial \theta}\right\rangle^2
+ \left\langle\frac{\partial ^2 H}{\partial \theta^2}\right\rangle\\
-\frac{1}{T} \left\langle\left(\frac{\partial H}{\partial \theta}\right)^2\right\rangle\\
  \label{ddf}
\end{equation} 
where the first term on the right hand side vanishes due to symmetry.
Evaluating the derivatives of $H$ for the Hamiltonian (\ref{VecHamiltonian}) gives the spin-wave stiffness
 $\rho_s=L^2 ({\partial ^2 f}/{\partial \theta^2})\big|_{\theta=0} $ as
\begin{equation}
\begin{gathered}
 \rho_s^\perp=\left\langle\sum_{\langle\mathbf r,\mathbf r'\rangle}J_{\mathbf r, \mathbf r'}\left[\mathbf S_{\mathbf r}\cdot\mathbf S_{\mathbf r'}-
(\mathbf S_{\mathbf r}\cdot \hat {\mathbf k})(\mathbf S_{\mathbf r'}\cdot \hat {\mathbf k})\right](z-z')^2 \right\rangle\\
-\frac{1}{T}\left\langle\left(\sum_{\langle\mathbf r,\mathbf r'\rangle}J_{\mathbf r, \mathbf r'}\left[(\mathbf S_{\mathbf r}\times \mathbf S_{\mathbf r'})\cdot
 \hat {\mathbf k}\right](z-z')\right)^2\right\rangle.
\label{rhoxy}
\end{gathered}
\end{equation}

Here, $\hat{\mathbf{k}}$ is the unit vector in the $z$-direction. The same equation was derived in Ref. \onlinecite{CaffarelAzariaDelamotteMouhanna94} for the $XY$ case.
Equation \ref{rhoxy} needs to be evaluated with fixed boundary conditions at the top and bottom layeres. Applying this formula to simulations
with periodic boundary conditions leads to incorrect results in the Heisenberg case (even though it works in $XY$ case).
The reason is that Eq. (\ref{rhoxy}) is sensitive to twist in the $XY$ plane only.
 
In the Heisenberg case this can be fixed by
 aligning the imaginary twist axis with a direction $\hat {\mathbf a}$ perpendicular to the total magnetization in each Monte-Carlo
 measurement. We use $\hat {\mathbf a}=(\mathbf m\times \hat {\mathbf k})/|\mathbf m\times \hat {\mathbf k}|$. 
 The resulting formula for the spin-wave stiffness can be used efficiently by Monte-Carlo
simulations with periodic boundary conditions. It reads
 \begin{equation}
\begin{gathered}
 \rho_s^\perp=\left\langle\sum_{\langle\mathbf r,\mathbf r'\rangle}J_{\mathbf r, \mathbf r'}\left[\mathbf S_{\mathbf r}\cdot\mathbf S_{\mathbf r'}-
(\mathbf S_{\mathbf r}\cdot \hat {\mathbf a})(\mathbf S_{\mathbf r'}\cdot \hat {\mathbf a})\right](z-z')^2 \right\rangle\\
-\frac{1}{T}\left\langle\left(\sum_{\langle\mathbf r,\mathbf r'\rangle}J_{\mathbf r, \mathbf r'}\left[(\mathbf S_{\mathbf r}\times \mathbf S_{\mathbf r'})\cdot
 \hat {\mathbf a}\right](z-z')\right)^2\right\rangle.
\label{rhoHeisenberg}
\end{gathered}
\end{equation}

We have tested that this equation reproduces the results obtained directly from Eq. (\ref{deltaf}).  

\bibliographystyle{apsrev4-1}
\bibliography{rareregions}

\end{document}